\documentclass{iopart}

\usepackage{url}
\usepackage{amsbsy}
\usepackage{graphicx}

\newcommand{\eqref}[1]{{(\ref{#1})}}
 
\begin{document}

\title{Report on the second Mock LISA Data Challenge} 

\author{The \emph{Mock LISA Data Challenge Task Force}:
Stanislav Babak$^1$,
John G. Baker$^2$,
Matthew J. Benacquista$^3$,
Neil J. Cornish$^4$,
Jeff Crowder$^5$,
Curt Cutler$^{5,6}$,
Shane L. Larson$^7$,
Tyson B. Littenberg$^4$,
Edward K. Porter$^1$,
Michele Vallisneri$^{5,6}$,
Alberto Vecchio$^{8,9}$ and \emph{the Challenge-2 participants}:
Gerard Auger$^{10}$, Leor Barack$^{11}$, Arkadiusz B\l aut$^{12}$, Ed Bloomer$^{13}$, Duncan A. Brown$^{14,15,6}$, Nelson Christensen$^{16}$, James Clark$^{13}$, Stephen Fairhurst$^{17,15,6}$, Jonathan R. Gair$^{18}$, Hubert Halloin$^{10}$, Martin Hendry$^{13}$, Arturo Jimenez$^3$, Andrzej Kr\'olak$^{19}$, Ilya Mandel$^{9,6}$, Chris Messenger$^{13}$, Renate Meyer$^{20}$, Soumya Mohanty$^3$, Rajesh Nayak$^3$, Antoine Petiteau$^{10}$, Matt Pitkin$^{13}$, Eric Plagnol$^{10}$, Reinhard Prix$^1$, Emma L. Robinson$^8$, Christian Roever$^{20}$, Pavlin Savov$^{6}$, Alexander Stroeer$^{8,9}$, Jennifer Toher$^{13}$, John Veitch$^8$, Jean--Yves Vinet$^{21}$, Linqing Wen$^1$, John T. Whelan$^1$, Graham Woan$^{13}$}
\address{$^1$ Max-Planck-Institut f\"ur Gravitationsphysik (Albert-Einstein-Institut), Am M\"uhlenberg 1, D-14476 Golm bei Potsdam, Germany}
\address{$^2$ Gravitational Astrophysics Lab., NASA Goddard Space Flight Center, 8800 Greenbelt Rd., Greenbelt, MD 20771, USA}
\address{$^3$ Center for Gravitational Wave Astronomy, Univ.\ of Texas at Brownsville, Brownsville, TX 78520, USA}
\address{$^4$ Dept.\ of Physics, Montana State Univ., Bozeman, MT 59717, USA}
\address{$^5$ Jet Propulsion Laboratory, California Inst.\ of Technology, Pasadena, CA 91109, USA}
\address{$^6$ Theoretical Astrophysics, California Inst.\ of Technology, Pasadena, CA 91125}
\address{$^7$ Dept.\ of Physics, Weber State Univ., 2508 University Circle, Ogden, UT 84408, USA}
\address{$^8$ School of Physics and Astronomy, Univ.\ of Birmingham, Edgbaston, Birmingham B152TT, UK}
\address{$^9$ Dept.\ of Physics and Astronomy, Northwestern Univ., Evanston, IL, USA}
\address{$^{10}$ APC, UMR 7164, Univ.\ Paris 7 Denis Diderot, 10, rue Alice Domon et Leonie Duquet, 75025 Paris Cedex 13, France}
\address{$^{11}$ School of Mathematics, Univ.\ of Southampton, Southampton, SO171BJ, UK}
\address{$^{12}$ Inst.\ of Theoretical Physics, Univ.\ of Wroc\l aw, Wroc\l aw, Poland}
\address{$^{13}$ Dept.\ of Physics and Astronomy, Univ.\ of Glasgow, Glasgow, UK}
\address{$^{14}$ Dept.\ of Physics, Syracuse Univ., Syracuse, NY13244, USA}
\address{$^{15}$ LIGO Laboratory, California Inst.\ of Technology, Pasadena, CA 91125}
\address{$^{16}$ Physics and Astronomy, Carleton College, Northfield, MN, USA}
\address{$^{17}$ School of Physics and Astronomy, Cardiff Univ., 5, TheParade, Cardiff, UK, CF243YB}
\address{$^{18}$ Inst.\ of Astronomy, Univ.\ of Cambridge, Cambridge, CB30HA, UK}
\address{$^{19}$ Inst.\ of Mathematics, Polish Academy of Sciences, Warsaw, Poland}
\address{$^{20}$ Dept.\ of Statistics, The Univ.\ of Auckland, Auckland, New Zealand}
\address{$^{21}$ ARTEMIS, Observatoire de la Cote d'Azur-C.N.R.S., 06304 Nice, France} 

\ead{Michele.Vallisneri@jpl.nasa.gov}

\begin{abstract}
The Mock LISA Data Challenges are a program to demonstrate LISA data-analysis capabilities and to encourage their development. Each round of challenges consists of several data sets containing simulated instrument noise and gravitational-wave sources of undisclosed parameters. Participants are asked to analyze the data sets and report the maximum information about source parameters. The challenges are being released in rounds of increasing complexity and realism: in this proceeding we present the results of Challenge 2, issued in January 2007, which successfully demonstrated the recovery of signals from supermassive black-hole binaries, from ~20,000 overlapping Galactic white-dwarf binaries, and from the extreme--mass-ratio inspirals of compact objects into central galactic black holes.
\end{abstract}

\vspace{-12pt}


\section{Introduction}

The Laser Interferometer Space Antenna (LISA), a NASA and ESA space mission to detect gravitational waves (GWs) in the $10^{-5}$--$10^{-1}$ Hz range \cite{lisa}, will produce time series consisting of the superposition of the signals from millions of sources, from our Galaxy to the edge of the observable universe. Some of the signals (such as those from extreme--mass-ratio inspirals, or EMRIs) are very complex functions of the physical parameters of the sources; others (such as those from Galactic white-dwarf binaries) are simpler, but their resolution will be confused by the presence of many other similar signals overlapping in frequency space. Thus, data analysis is integral to the LISA measurement concept, because no source can be identified without first carefully teasing out its individual voice from the noisy party of each data set. Understanding data analysis is therefore important to demonstrate that LISA can meet its science requirements, and to translate them into decisions about instrument design.

The idea of the Mock LISA Data Challenges (MLDCs) arose in late 2005 from this realization. The MLDCs have the purpose of encouraging and tracking progress in LISA data-analysis development, and (as a useful byproduct) of prototyping the LISA computational infrastructure: common data formats, standard models of the LISA orbits, noises and measurements, software to generate waveforms and to simulate the LISA response, and more.
The MLDCs are a coordinated (but voluntary) effort in the GW community, whereby a task force chartered by the LISA International Science Team periodically issues a number of data sets containing 
synthetic noise and GW signals from sources of undisclosed parameters; challenge participants return detection candidates and parameter estimates, together with descriptions of their search methods. These results are then compiled and compared to the previously secret challenge ``key''.

Challenge 1, issued in Jun 2006 with results due in Dec 2006 (see \cite{mldclisasymp,mldcgwdaw1}), tackled the detection and parameter characterization of \emph{verification binaries} (Galactic binaries of known frequency and position); of loud unknown Galactic binaries, either alone or in small, moderately interfering groups; and of relatively loud inspirals of supermassive--black-hole (MBH)  binaries. All sources were represented by somewhat idealized waveforms, and they were staged on instrument noise alone. Altogether, Challenge 1 successfully demonstrated the detection of all three source classes. Ten research collaborations submitted entries, adopting a variety of methods (template-bank, stochastic and genetic matched filtering; time--frequency; tomography; Hilbert transform). Despite the short timescale, all challenges were ``solved'' by at least one group, although some searches locked on strong secondary maxima of the source parameter probabilities. More important, Challenge 1 helped set the playing field and assemble the computational tools for the more realistic Challenge 2.

Challenge 2, issued in Jan 2007 with results due at the end of Jun 2007, raised the bar by proposing three complex subchallenges. Data set 2.1 contained a full population of Galactic binary systems (about 26 million sources). Data set 2.2 contained a different realization of the Galaxy, plus 4--6 MBH binary inspirals with single-interferometer signal-to-noise ratios (SNRs) between 10 and 2000 and a variety of coalescence times, and five EMRIs with SNRs between 30 and 100. Last, five more data sets (denoted 1.3.1--5, since they were actually released at the time of Challenge 1) contained a single EMRI signal over instrument noise alone. See \cite{mldcgwdaw2} for more details about the signal models and the exact source content of the data sets.

Thirteen collaborations (comprising all the researchers listed as participants in the byline of this article, and most task force members) submitted a total of 22 entries, including a proof-of-principle analysis for stochastic backgrounds performed on data set 2.1.
Altogether, Challenge 2 successfully demonstrated the identification of $\sim 20,000$ Galactic binaries, the accurate estimation of MBH inspiral parameters, and the positive detection of EMRIs. In the rest of this paper, we describe some of its highlights. All the solutions submitted by participating groups, together with technical write-ups of their methods and findings, can be found at the URL \url{www.tapir.caltech.edu/~mldc/results2}. A few groups are also contributing descriptions of their work to the proceedings of this conference.

\section{Data sets 2.1 and 2.2: The Galaxy}

Five groups submitted Galactic-binary catalogs for data sets 2.1 and 2.2:
\begin{description}
\item[GLIG] A collaboration of research groups at institutions in the UK, United States and New Zealand developed a Reversible-Jump Markov Chain Monte Carlo (RJ MCMC) code that can sample models with different numbers of sources; for lack of time, however, they only submitted parameter sets for the verification binaries.
\item[IMPAN] Kr\'olak and Blaut developed an $\mathcal{F}$-statistic, template-bank--based matched-filtering search \cite{JKS98,KTV04}, and submitted parameters for 404 sources for data set 2.1.
\item[MTJPL] The Montana State--JPL collaboration used a Metropolis--Hastings Monte Carlo (MHMC) code that ran separately for overlapping frequency bands and for different hypotesized numbers of sources; model comparison was then used to determine the most probable number of sources in each band. The collaboration submitted parameter sets for 19,324 sources for data set 2.1, and 18,461 sources for data set 2.2.
\item[PrixWhelanAEI] Prix and Whelan developed an $\mathcal{F}$-statistic,
template-bank--based matched-filtering search using a hierarchical
scheme that enforced trigger coincidence between TDI observables,
followed by a coherent follow-up using noise-orthogonal TDI
combinations \cite{prixwhelan}. They submitted parameter sets for 1777 sources for data set 2.1, and 1737 sources for data set 2.2.
\item[UTB] Nayak, Jimenez and Mohanty used a tomographic reconstruction technique and submitted parameters for 3862 sources in data set 2.1.
\end{description}
Evaluating the performance of these searches brings up several problems of principle: while we know that many of the $\sim 30$ million Galactic sources that were injected into the data sets cannot be recovered because they are (relatively) too weak, we do not have a precise estimate of how many sources should be recoverable. Thus, the notion of false dismissal is not well defined. To make matters worse, the notion of false positive is also ill defined, because a single recovered source can provide a good fit to the blended signal from several injected sources, which may well be the ``right'' answer with the knowledge we have, since it is the best fit to the data with the smallest numbers of parameters.

The task force devoted considerable time to the analysis of Galactic-binary searches, and we do not have space here to describe all the treatments that we applied to the data. Instead we will limit our report to techniques that pair up individual recovered sources with individual sources from the challenge key, with the understanding that this will overestimate the number of false positives, and say nothing about false dismissals.
\begin{figure}
\includegraphics[width=\textwidth]{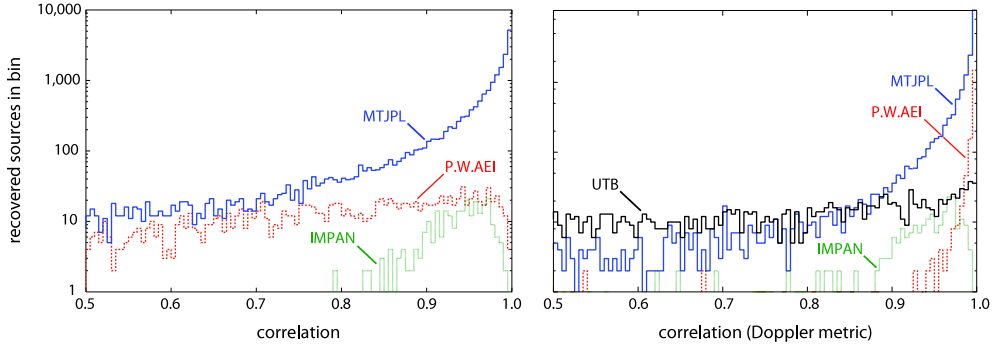}
\caption{Correlation analysis of Challenge-2.1 Galactic-binary catalogs (histogram, with bin fractions given on logarithmic scale).  Left panel: associated by correlation; right panel: associated by Doppler metric.\label{fig:correlation}}
\end{figure}

One way to proceed is to associate the reported and injected sources that have the strongest signal correlation, limiting the search to the bright ($\mathrm{SNR} > 2$) injected sources that could in principle have been found: in the left panel of figure \ref{fig:correlation} we show the distribution of correlations generated with this procedure for data set 2.1.
Detections with the highest correlations can be considered ``safest,'' while those with the lowest correlations probably represent spurious associations.

Another procedure is to associate the reported and injected sources that minimize the \emph{Doppler metric} that spans the frequency--sky-location subspace of the full parameter space, and automatically maximizes correlation over the \emph{extrinsic parameters} (amplitude, polarization, inclination, initial phase): the right panel of figure \ref{fig:correlation} shows the resulting distribution of correlations. The UTB entry, which includes frequency and sky position but not the extrinsic parameters, can only be plotted this way. Generally, this is a softer criterion, and all searches do better by it (especially the PrixWhelanAEI entry, whose long-wavelength approximation for the LISA response is prone to extrinsic-parameter errors).
\begin{figure}
\includegraphics[width=\textwidth]{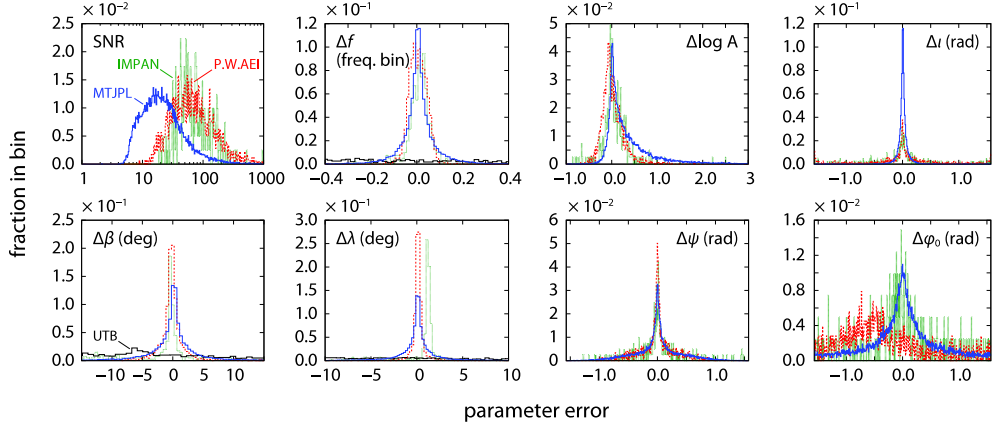}
\caption{Recovered SNRs and intrinsic and extrinsic parameter errors for Challenge-2.1 Galactic-binary catalogs (histogram). True sources and templates are associated by correlation, except for the UTB catalog, for which they are associated by Doppler metric.\label{fig:paramerrors}}
\end{figure}

Figure \ref{fig:paramerrors} shows the SNRs of the recovered sources and the errors for the intrinsic and extrinsic parameters, computed after associating sources by correlation (and by Doppler metric only for the UTB entry), again for data set 2.1. The error in frequency is in most cases within a small fraction of a Fourier bin, and the errors in sky position are within a few degrees; by contrast, the errors in the amplitude and in the (extrinsic) orientation angles are larger. The $\phi_0$ graph for the PrixWhelanAEI suggests a systematic error in the definition of initial phase.

Altogether, these challenges demonstrated a solid capability in analyzing signals from the Galaxy and resolving a large number of binaries. As we mentioned, deciding how well they were recovered is not an easy question to answer, because of the difficulty of defining (at least operationally) a notion of \emph{identity} for recovered sources. These problems deserve careful attention in the future.

\section{Data set 2.2: MBH binaries (over the Galaxy)}

Four groups reported parameter sets for the MBH binaries in data set 2.2:
\begin{description} 
\item[AEIse] Babak and Porter used an $\mathcal{F}$-statistic, template-bank--based matched-filtering search, followed by an MCMC stage.
\item[MTAEI] Cornish and Porter used an MHMC matched-filtering search with a \emph{frequency-annealed} scheme where shorter, lower-frequency templates are used in the initial phases of the search and then progressively extended.
\item[JPLCT] The JPL--CIT collaboration used a three-stage pipeline consisting of a track search in the time--frequency (TF) plane, followed by template-bank--based matched filtering, and by an MCMC refinement \cite{brown}.
\item[LisaFrance] The French collaboration used a TF track search alone, and therefore reported only mass and time-of-coalescence parameters.
\end{description}
All the four MBH binaries in the data set (MBH-1, 2, 4 and 5, with total SNRs $\sim$ 2583, 25, 174 and 117) were positively detected by AEIse, MTAEI, and JPLCT; the TF method used by LisaFrance identified MBH-1 and 4, but not MBH-2, and could report only a time of coalescence for MBH-5.
\begin{table}
\caption{\label{tab:mbh1}%
Recovered SNR and parameter errors for MBH-1. All parameters are defined as in tables 2 and 4 of \cite{mldcgwdaw2}, except for $\mu = m_1 m_2 / (m_1 + m_2)$, $M_c = (m_1 m_2)^{3/5} / (m_1 + m_2)^{1/5}$, and $\phi_c$ defined as the GW phase at coalescence. All errors on angles are given in radians; the true (optimal) SNR is \textbf{2583.42}.}
\small%
\begin{tabular}{@{}l|r@{\;}r@{\;}r@{\;}r@{\;}r@{\;}r@{\;}r@{\;}r@{\;}r@{\;}r}
\br
           & SNR & $\Delta M_c/M_c$ & $\Delta \mu/\mu$ & $\Delta t_c/t_c$ & $\Delta \beta$ & $\Delta \lambda$ & $\Delta D / D$ & $\Delta \iota$ & $\Delta \psi$ & $\Delta \phi_c$ \\
           &     & $\times 10^{-5}$ & $\times 10^{-5}$ & $\times 10^{-6}$ & $\times 10^{-1}$ & $\times 10^{-2}$ & $\times 10^{-2}$ & $\times 10^{-2}$ & $\times 10^{-1}$ & $\times 10^{-1}$ \\
\mr
AEIse      & 2247.60 &  147.0 & 3386.2 & 19.0 & $-5.07$ & $-82.1$ & 77.8 & $-6.86$ &  13.8 & $-6.82$ \\
MTAEI      & 2583.34 &    7.9 &    9.1 &  3.6 & 1.65 &  $-1.2$ &  3.5 &  4.94 &   1.2 & $-7.70$ \\
JPLCT      & 2582.42 &   27.5 &   28.7 & 16.0 & 4.81 &  12.2 & 12.3 & $-3.19$ & $-12.1$ &  7.45 \\
lisaFrance & \multicolumn{1}{c}{--} & 2944.6 &   72.7 & 67.8 & \multicolumn{1}{c}{--}    &  \multicolumn{1}{c}{--}   & \multicolumn{1}{c}{--}   & \multicolumn{1}{c}{--}    & \multicolumn{1}{c}{--}    &  \multicolumn{1}{c}{--}   \\
\br
\end{tabular}
\end{table}
\begin{table}
\caption{Recovered SNR and parameter errors for MBH-4, given as in table \ref{tab:mbh1}. The true (optimal) SNR is \textbf{174.12}.\label{tab:mbh4}}
\small%
\begin{tabular}{@{}l|r@{\;}r@{\;}r@{\;}r@{\;}r@{\;}r@{\;}r@{\;}r@{\;}r@{\;}r}
\br
           & SNR & $\Delta M_c/M_c$ & $\Delta \mu/\mu$ & $\Delta t_c/t_c$ & $\Delta \beta$ & $\Delta \lambda$ & $\Delta D / D$ & $\Delta \iota$ & $\Delta \psi$ & $\Delta \phi_c$ \\
           &     & $\times 10^{-6}$ & $\times 10^{-4}$ & $\times 10^{-6}$ & $\times 10^{-2}$ & $\times 10^{-2}$ & $\times 10^{-3}$ & $\times 10^{-3}$ & $\times 10^{-3}$ & $\times 10^{-1}$ \\
\mr
AEIse       & 81.38     &   1396.3  &  149.9 &   3.4 & $-12.5$ & $ 104.7$ &  574.0 &  $  3.5$ &   $-185.4$ & $  8.1$ \\
MTAEI       & 174.13    &    148.8  &   21.3 &   2.1 & $2.4$ & $   2.1$ &   15.1 &  $  2.8$ &   $  -7.5$ & $  1.7$ \\
            & 174.11    &     17.1  &   20.5 &  33.3 & $-42.4$ & $-310.6$ &   16.7 &  $-13.7$ &   $-146.3$ & $ -6.3$ \\
JPLCT       & 174.11    &      4.2  &    9.5 &   2.1 & $7.8$ & $   9.6$ &    1.2 &  $  1.3$ &   $ -21.3$ & $ -5.4$ \\
            & 174.12    &    124.7  &    9.0 &  35.4 & $-47.3$ & $-302.9$ &    6.3 &  $-12.4$ &   $1436.4$ & $-12.4$ \\
lisaFrance  & \multicolumn{1}{c}{--}        &  34394.1  & 1804.1 & 280.8 &  \multicolumn{1}{c}{--}    &  \multicolumn{1}{c}{--}      &  \multicolumn{1}{c}{--}    &    \multicolumn{1}{c}{--}   &   \multicolumn{1}{c}{--}       & \multicolumn{1}{c}{--}      \\
\br
\end{tabular}
\end{table}

Tables \ref{tab:mbh1} and \ref{tab:mbh4} show fractional parameter errors for MBH-1 and 4, together with the SNR recovered by the best-fit candidates, computed as 
\begin{equation}
\mathrm{SNR}_\mathrm{best}  = \frac{(A_{true}|A_{best}) + (E_{true}|E_{best})}
{\sqrt{(A_{best}|A_{best}) + (E_{best}|E_{best})}},
\end{equation}
with $(\cdot|\cdot)$ the usual noise-weighted inner product, and $A = (2X - Y - Z)/3$ and $E = (Z-Y)/\sqrt{3}$ two noise-orthogonal TDI observables (see, e.g., \cite{VCT}). In table \ref{tab:mbh1}, we see that the JPLCT search for MBH-1 locked onto a secondary probability maximum with $\mathrm{SNR}_\mathrm{best}$ only slightly lower than the optimal value, but with sky positions off by several degrees, which also led to errors in the other parameters. The JPLCT authors hypothesize that this was caused by first subtracting a rough MBH-1 model from the data, then subtracting the resolvable Galactic binaries, and finally refining the MBH search.

MBH-4 (table \ref{tab:mbh1}) is an interesting example of a ``true'' bimodal probability distribution for the source parameters. MTAEI and JPLCT each submitted two candidates, placed at rather different sky locations, quoting relative probability ratios of 1:1 and 1.18:1. In this case, it was probably the sky position and orientation of this source that conspired to degrade LISA's positional sensitivity, since they resulted in a very weak signal in one of the noise-orthogonal observables.

Altogether, this challenge demonstrated a solid capability in the detection and parameter estimation of MBH inspirals with moderate SNRs, even in the presence of a strong Galactic background, at least if the inspirals can be considered close to our idealized model: circular and adiabatic with negligible spin effects. These restrictions are being relaxed for the upcoming Challenge 3.

\section{Data sets 1.3.X: EMRIs}

Three groups reported parameter sets for the EMRIs in data sets 1.3.1--1.3.4. No group tackled the problem of detecting these systems in data set 2.2 (on top of the Galactic background).
\begin{description}
\item[BBGP] Babak and colleagues used an MCMC matched-filtering search that modeled the signal with a sequence of progressively longer templates (a \emph{time-annealed} scheme).
\item[EtfAG] Gair, Mandel and Wen used a TF track search that (for now) targeted only the intrinsic parameters and sky position \cite{gair}.
\item[MT] Cornish used an MHMC matched-filtering search, running it in parallel on individual month-long segments, which were subsequently strung together for full detections.
\end{description}
Table \ref{tab:emri} shows typical recovered SNRs and errors. While it is clear that the matched-filtering searches locked on several secondary probability maxima with comparable probabilities, the recovered SNRs correspond to solid detections with exceedingly low false-alarm probabilities. The errors are quoted as fractions of the allowed parameter ranges, and they are quite large. Intriguingly, the TF search was the most accurate in determining the sky position. Altogether, these challenges demonstrated a positive capability of detecting EMRIs, at least if their signals are similar in complexity to the \emph{kludge} waveforms used in this challenge \cite{mldcgwdaw2}; however, the prospects for accurate parameter estimation are still uncertain, and a good focus for further challenges.
\begin{table}
\caption{Recovered SNRs and parameter errors for the EMRI signal in data set 1.3.1. All errors are given as \emph{fractions of the allowed prior range} for the corresponding parameters (0.15 for $e_0$), except for the errors on $\nu_0$ and $D$. Not all parameters are shown. For their definitions, see tables 2 and 5 of \cite{mldcgwdaw2}. The true (optimal) SNR is \textbf{130.98}.\label{tab:emri}}
\small%
\lineup
\begin{tabular}{@{}l@{\;}|@{\;}l@{\;}@{\;}l@{\;}l@{\;}l@{\;}l@{\;}l@{\;}l@{\;}l@{\;}l@{\;}l@{\;}l@{\;}l@{}}
\br
& SNR & \multicolumn{1}{c}{$\delta \beta$} & \multicolumn{1}{c}{$\delta \lambda$} & \multicolumn{1}{c}{$\delta \theta_K$} & \multicolumn{1}{c}{$\delta \phi_K$} & \multicolumn{1}{c}{$\delta a$} & \multicolumn{1}{c}{$\delta \mu$} & \multicolumn{1}{c}{$\delta M$} & \multicolumn{1}{c}{$\frac{\Delta \nu_0}{\nu_0}$} & \multicolumn{1}{c}{$\delta e_0$} & \multicolumn{1}{c}{$\frac{\Delta D}{D}$} \\
\mr
BBGP    & 74.86  & $-0.33 $   & $-0.0095$   & $-0.13 $ & $-0.076$ & $\m 0.28 $ & $-0.15$   & $-0.51$   & $\m 0.017  $ & $\m 0.21 $ &  $-1.21$ \\
        & 72.96  & $-0.32 $   & $\m 0.011 $ & $-0.15 $ & $-0.078$ & $\m 0.27 $ & $-0.15$   & $-0.51$   & $\m 0.017  $ & $\m 0.21 $ &  $-1.22$ \\
        & 72.52  & $-0.28 $   & $\m 0.025 $ & $-0.063$ & $-0.036$ & $\m 0.41 $ & $-0.17$   & $-0.35$   & $-0.009  $   & $\m 0.29 $ &  $-2.15$ \\
        & 72.49  & $-0.28 $   & $\m 0.025 $ & $-0.063$ & $-0.034$ & $\m 0.41 $ & $-0.17$   & $-0.36$   & $-0.009  $   & $\m 0.29 $ &  $-2.17$ \\
        & 70.59  & $-0.31 $   & $-0.020 $   & $-0.36 $ & $-0.21 $ & $\m 0.44 $ & $-0.12$   & $-0.12$   & $-0.03   $   & $\m 0.28 $ &  $-0.91$ \\
EtfAG   & \multicolumn{1}{c}{--}  & $\m 0.016$ & $\m 0.0012$ & \multicolumn{1}{c}{--}    & \multicolumn{1}{c}{--}        & $-0.082$   & $\m 0.10 $ & $-0.17$   & $\m 0.0026 $ &  $\m 0.098$ &   \multicolumn{1}{c}{--}   \\
MT      & 74.85 & $\m 0.15 $ & $\m 0.47  $ & $-0.069$ & $-0.15 $ & $-0.026$   & $\m 0.073$ & $\m 0.18$ & $\m 0.00025$  & $-0.11 $   &  $-0.71$ \\
        & 76.52  & $\m 0.084$ & $-0.49  $   & $-0.33 $ & $-0.10 $ & $-0.022$   & $\m 0.046$ & $\m 0.16$ & $\m 0.00026$ & $-0.10 $   &  $-0.70$ \\
\br
\end{tabular}
\end{table}

\section{Conclusion}

We are very excited about the outcome of the first two MLDCs, which have given a convincing demonstration that a significant portion of the LISA science objectives could already be achieved with techniques that are currently in hand. Most of the research groups that participated in Challenge 1 have successfully made the transition to the greater complexity of Challenge 2. Challenge 3 will continue to move in the direction of more realistic signals, featuring chirping Galactic binaries and precessing binaries of spinning MBHs. It will also include two new classes of signals: an isotropic primordial GW background and bursts from the cusps of cosmic strings. Between July and Dec 2007 we are also running Challenge 1B, a repeat of Challenge 1 conceived to provide a softer entry point for research groups new to the MLDCs.

The MLDC conventions, file formats, and software tools (see \url{lisatools.googlecode.com}) have matured to the point where interested parties can use them to generate a variety of data sets. This enables a wealth of interesting side investigations, such as the studies of the LISA science reach that are now being undertaken by the LISA Science Team. To obtain more information and to participate in the MLCDs, see the official MLDC website (\url{astrogravs.nasa.gov/docs/mldc}) and the task force wiki (\url{www.tapir.caltech.edu/listwg1b}).


\ack

SB, EKP, and JTW's work was supported by the German Aerospace Center (DLR) and the Max-Planck Society. MB acknowledges funding from NASA Grant NNG04GD52G and support from the NASA Center for Gravitational Wave Astronomy at University of Texas at Brownsville (NAG5-13396). NC was supported by NASA Grants NNG05GI69G and NNX07AJ61G. DB and SF acknowledge funding from NSF grant PHY-0601459 and the LIGO Laboratory. JC's, CC's and MV's work was carried out at the Jet Propulsion Laboratory, California Institute of Technology, under contract with the National Aeronautics and Space Administration. JG acknowledges support from St Catharine's College, Cambridge. IM would like to thank the Brinson Foundation, NASA grant NNG04GK98G and NSF grant PHY-0601459 for financial support. RP's work was supported by the Max-Planck Society. MV was supported by LISA Mission Science Office and by JPL's Human Resources Development Fund. LW's work was supported by the Alexander von Humboldt Foundation's Sofja Kovalevskaja Programme, funded by the German Federal Ministry of Education and Research.




\section*{References}

\begin{thebibliography}{99}
%
\bibitem{lisa} Bender P, Danzmann P and
  the LISA Study Team 1998 ``Laser Interferometer Space Antenna for the Detection of Gravitational Waves, Pre-Phase A Report'' \textbf{MPQ 233} (Garching: Max-Planck-Instit\"ut f\"ur
  Quantenoptik) 
%
\bibitem{mldclisasymp} Arnaud K A et al. (the Mock LISA Data Challenge Task Force) 2006 
\textit{Laser Interferometer Space Antenna: 6th International LISA Symp. (Greenbelt, MD, 19--23 Jun 2006)} ed Merkowitz S M and Livas J C (Melville, NY: AIP) p 619; \textit{ibid.} p 625
%
\bibitem{mldcgwdaw1} Arnaud K A et al. (the Mock LISA Data Challenge Task Force and Challenge 1 participants) 2007 \textit{Class. Quant. Grav.} \textbf{24} S529
%
\bibitem{mldcgwdaw2} Arnaud K A et al. (the Mock LISA Data Challenge Task Force) 2007 \textit{Class. Quant. Grav.} \textbf{24} S551
%
\bibitem{JKS98}
P.\ Jaranowski, A.\ Kr\'olak, and B.\ F.\ Schutz, Phys.\ Rev.\ D
{\textbf 58}, 063001 (1998).
%
\bibitem{KTV04} A. Kr\'olak, M. Tinto, and M. Vallisneri, {\it Phys. Rev. D}, {\bf 70},
022003 (2004).
%
\bibitem{jpcs} Various authors 2008 Proc. 7th Amaldi Conf. on Gravitational Waves (Sydney, 8--14 July 2007) \textit{J. Phys. Conf. Ser.} \textbf{XX}
%
\bibitem{brown} Brown D A, Crowder J, Cutler C, Mandel I and Vallisneri M 2007 \textit{Class. Quant. Grav.} \textbf{24} S595
%
\bibitem{VCT} Vallisneri M, Crowder J and Tinto M 2007 ``Sensitivity and parameter-estimation precision for alternate LISA configurations'' \textit{Preprint} arxiv.org/0710.4369 
%
\bibitem{gair} Gair J R, Mandel I and Wen L 2007 Proc. 7th Amaldi Conf. on Gravitational Waves (Sydney, 8--14 July 2007), submitted. \textit{Preprint} arXiv:0710.5250
%
\bibitem{prixwhelan} Prix R and Whelan J T 2007 \textit{Class. Quant. Grav.} \textbf{24} S565; \textit{Poster} \url{www.ligo.caltech.edu/docs/G/G070462-00.pdf}
%
\end{thebibliography}
\end{document}